\documentclass{article}
\usepackage{spconf,amsmath,graphicx,amssymb}

\usepackage{multirow}
\usepackage{color}
\usepackage{subcaption}
\usepackage{hyperref}

\usepackage{xcolor}

\definecolor{pink}{RGB}{219, 48, 122}

\usepackage[numbers,sort&compress]{natbib}

\usepackage{bm}
\usepackage{amssymb,amsmath,graphicx,bbm,color,booktabs}
\usepackage{amsopn}

    \newcommand{\Cc}{\mathcal{C}}

\newcommand{\R}{\mathbb{R}}

\newcommand{\X}{\mathcal{X}}

\renewcommand{\eqref}[1]{Eq.~(\ref{#1})}

\newcommand{\audiogen}{\textsc{AudioGen}}

\def \x{{\mathbf x}}
\def \c{{\mathbf c}}

\title{Audio Language Modeling using \\ Perceptually-Guided Discrete Representations}
\name{\begin{tabular}{c}
    \it  Felix Kreuk$^{1}$, Yaniv Taigman$^{1}$, Adam Polyak$^{1}$, Jade Copet$^{1}$,\\
    \it Gabriel Synnaeve$^{1}$, Alexandre D\'efossez$^{1}$, Yossi Adi$^{1,2}$
\end{tabular}}
\address{$^1$Meta AI Research
\\
$^2$The Hebrew University of Jerusalem, Israel}

\begin{document}

\maketitle

\begin{abstract}
In this work, we study the task of Audio Language Modeling, in which we aim at learning probabilistic models for audio that can be used for generation and completion. We use a state-of-the-art perceptually-guided audio compression model, to encode audio to discrete representations. Next, we train a transformer-based causal language model using these representations. At inference time, we perform audio auto-completion by encoding an audio prompt as a discrete sequence, feeding it to the audio language model, sampling from the model, and synthesizing the corresponding time-domain signal. We evaluate the quality of samples generated by our method on Audioset, the largest dataset for general audio to date, and show that it is superior to the evaluated baseline audio encoders. We additionally provide an extensive analysis to better understand the trade-off between audio-quality and language-modeling capabilities. Samples: \href{https://adiyoss.github.io/audio-cont/}{link}.
\end{abstract}
\noindent\textbf{Index Terms}: Audio synthesis, Language Modeling

\section{Introduction}
Audio Language Modeling is the task of learning probabilistic models for audio, such models can be used for generation~\cite{van2017neural}, completion~\cite{lakhotia2021generative, kharitonov2021text}, classification~\cite{zhang2017learning}, and compression~\cite{krishnamurthy2009audio}. Previous work has mainly focused on limited domains of audio such as speech~\cite{van2017neural, lakhotia2021generative, kharitonov2021text, kreuk2021textless, borsos2022audiolm} and music~\cite{dhariwal2020jukebox, borsos2022audiolm}. In this paper we explore the modeling of general audio coming from a wide variety of domains, with a focus on inference speed, model simplicity, and perceptually-guided losses.

Recently, speech synthesis has seen great advances, leading to high quality natural sounding speech. Specifically, in the field of Generative Spoken Language Modeling (GSLM), the authors of~\cite{lakhotia2021generative, kharitonov2021text, kreuk2021textless} have shown that speech signals can be represented as discrete sequences and modeled using language models, similarly to text. The discrete representation is obtained by training a Self-Supervised Learning (SSL) speech representation model, and applying k-means clustering where the cluster indices are the discrete variables representing the original recording. While showing great promise for speech modeling, this approach does not directly translate to general audio modeling for several reasons. The first being that while some representations are effective for speech, they are not suitable for general audio. For example, Mel-Frequency Cepstral Coefficients are used as supervision for the HuBERT~\cite{hsu2021hubert} model, however these are less suitable for general audio purposes. Second, most self-supervised representations for speech operate at a low sampling rate. This results in a compact representation, which is effective when the underlying signal is not rapidly changing (e.g., speech), but fails to represent rich signal like general audio. Concurrent to our work, the authors of AudioLM~\cite{borsos2022audiolm} achieved discretization via residual vector quantization \cite{zeghidour2021soundstream} and further improved long-term dependency modeling by introducing semantic audio tokens.

\begin{figure}[t!]
  \centering
  \includegraphics[width=0.75\linewidth]{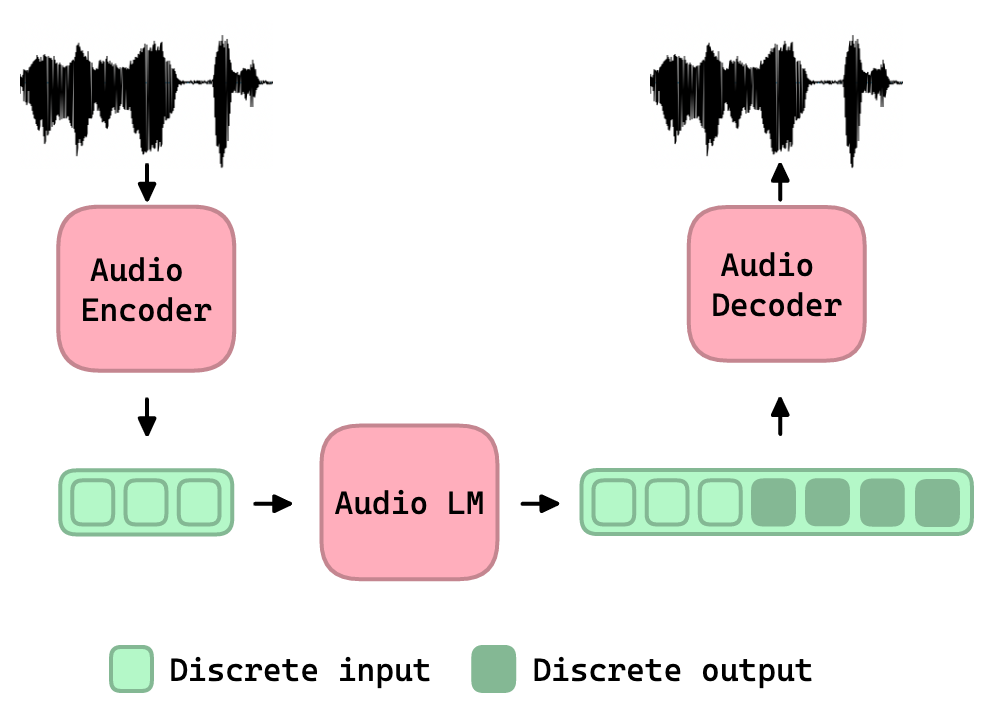}
  \caption{A general overview of the proposed system. The input is first being encoded into discrete sequence of units, next the Audio Language Model generates the continuation (marked as dark rectangles), lastly, the audio is being generated from the sequence of units.}
  \vspace{-0.3cm}
  \label{fig:arch}
\end{figure}

\audiogen~was recently proposed for text-to-audio generation~\cite{kreuk2022audiogen}. In this work, the authors proposed a model that perform textually guided audio generation together with audio continuation. First a codebook is being learned over the domain of audio samples, then a Transformer-decoder is being applied over the audio discrete representation. Unlike previous work, \audiogen~uses an additional text encoder to steer the generation process towards the textual description. 

In this work, we focus on audio continuation. Given a general audio input prompt, we generate an audio continuation. We follow a similar setup to \audiogen, while first training an audio compression network to encode input signals as discrete sequences sampled at a lower frequency. Then, using this network as a feature extractor we discretize a large collection of general audio waveforms, and train a Transformer-decoder causal language model. We empirically show that the proposed approach is able to generate general audio and outperforms existing baselines. We compare against other state-of-the-art audio encoders together with an analysis into what factors of an audio modeling system are important for high-quality audio generation.
\vspace{-0.1cm}
\section{Related work}
\label{sec:rel}
\vspace{-0.1cm}
Audio modeling is a notoriously challenging task due to its high-dimensionality (e.g., a single audio file sampled at 16kHz will contain 16,000 timesteps). When considering general audio the task is even harder due to a high variance in the input signal (unlike speech). Previous attempts can be roughly divided in two main categories: i) auto-regressive modeling over the raw waveform~\cite{chung2015recurrent,oord2016wavenet,child2019generating,goel2022s}; ii) representing audio as low bit-rate discrete representation and applying Natural Language Modeling techniques to model it~\cite{van2017neural,dhariwal2020jukebox,kharitonov2021text,lakhotia2021generative,kreuk2021textless,gat2022robustness,borsos2022audiolm}.

While models that operate on the raw waveform show success in subjective studies, they are mainly trained on specific or limited audio domains (e.g., musical instruments, read digits) only. Moreover, these often fail to capture linguistic structure when not conditioned on text~\cite{oord2016wavenet}. On the other hand, models working on low bit-rate discrete representation were mainly evaluated under low sampling rates (e.g., each discrete unit corresponds to 20ms from the raw-audio). This reduces the sequence length and results in a compact representation, which is suitable for spoken speech modeling~\cite{lakhotia2021generative,kharitonov2021text, nguyen2022generative}, emotion conversion~\cite{kreuk2021textless}, and speech-to-speech translation~\cite{lee2021direct,lee2021textless,popuri2022enhanced}, however fails to reconstruct rich signals. 

To maintain long-term consistency in music generation, the authors of~\cite{dhariwal2020jukebox} proposed to model music in a hierarchy of discrete representations. Practically, the authors train three separate VQ-VAE models for 1-dimensional signals using different hop-lengths. Then, using the quantized latent representation of each of these models, a separate language model was trained per quantization model. Sampling from this system involves with sampling from each language model in the hierarchy in a procedure called ancestral sampling. Finally, decoding the sampled sequence back to the time-domain is performed by using an auto-regressive WaveNet-like models resulting in costly inference.

Recently, the authors in~\cite{borsos2022audiolm} proposed the modeling speech and piano music using a set of acoustic and semantic tokens. Such modeling paradigm shows impressive results considering zero-resource speech challenge metrics~\cite{dunbar2021zero}. In the study, we focus on acoustic tokens for general audio synthesis (e.g., environmental sounds, fully mixed music). Specifically, we explore different SSL methods to create such tokens, and analyze the impact of temporal resolution and vocabulary size.

The authors in~\cite{kreuk2022audiogen} recently proposed a language modeling approach for audio generation conditioned on textual inputs. Although they provide impressive results, a detailed analysis on general audio continuation is still missing. Inspired by~\cite{kreuk2022audiogen, borsos2022audiolm}, in this work we analyze the performance of such model using acoustic tokens on audio continuation considering subjective evaluation, and provide a comparison to other audio encoders. 
\vspace{-0.2cm}
\section{Model}
\vspace{-0.1cm}
We follow a similar setup to \audiogen~\cite{kreuk2022audiogen}, which is composed of three main components: i) audio encoder; ii) audio decoder; and a iii) language model. We first train an audio auto-encoder model with a quantization bottleneck. Next, we use the discrete representations obtained from this model for learning a language model. Then, we perform conditional audio generation by priming the language model on a segment of audio, and generating a suitable continuation. See a visual depiction of the proposed system in Figure~\ref{fig:arch}.

Formally, we denote the domain of audio samples by $\X \subset \R$. The representation for a raw speech signal is therefore a sequence of samples $\x = (x_1,\ldots, x_T)$, where  $x_t\in\X$ for all $1\leq t \leq T$. The length of the input signal varies, thus the number of input samples in the sequence, $T$, is not fixed. We denote by $\X^T$ the set of all $T$-length sequences over $\X$.

\noindent {\bf Audio representation.}
To model audio we first encode it as a sequence of discrete units, with a temporal-resolution (hop-length) of $R$ milliseconds, and a vocabulary size of $K$. This transformation is performed using an audio compression model $A$, consisting of an encoder $E: \X^T \rightarrow \R^{T' \times d}$, a vector-quantization bottleneck $V: \R^{T' \times d} \rightarrow \Cc^{T'}$ where $\Cc=\{1, \dots, K\}$, that maps latent representations to indices of a learned codebook of size $K$. We denote $\c=(c_1,\dots,c_{T'})$, where $c_i \in \Cc$, as the output of applying $E \circ V$ over $\x$. Last, a decoder $D: \Cc^{T'} \rightarrow \X^T$ that reconstructs the time-domain signal from the quantized latent embeddings. We follow the same setting as in~\cite{defossez2022high, kreuk2022audiogen}, where the audio encoder is an auto-encoder trained to minimize a set of reconstruction losses, including L1 reconstruction loss; Multi-Resolution STFT loss~\cite{arik2018fast}; and Multi-Scale STFT discriminator loss. For vector quantization, the encoder minimizes the VQ-VAE objective \cite{van2017neural}, where the codebook is updated using an exponential moving average.

\noindent {\bf Audio language modeling.}
To correctly model audio one would ideally need to learn $p(\x)$. However, due to the high sampling rate of audio this becomes costly, and achieving temporal consistency over large samples becomes difficult.
Alternatively, given an audio encoder, one could instead learn the prior over the space of quantized latent representation.

Given an input waveform $\x$ we encode it as a sequence of discrete units as $\c = V(E(\x))$. Next, a transformer-decoder model is trained as a causal language model over the sequences of discrete units. Formally, given a training set of $n$ audio samples $\{\x_1,\dots,\x_n\}$, we first encode them into a discrete representation, $\{\c_1, \dots, \c_n \}$. Then, we optimize the cross-entropy loss function as follows, 
    $L = - \sum_{i=1}^n \sum_{j=1}^{T'} \log p(\c_i^j | \c_i^{j-1}, \dots, \c_i^0)$,
where $T'$ is the length of the encoded sequence. 

Empirically, we see a trade-off between audio-quality and language modeling capabilities, controlled by $R$ and $K$. As the encoder operates at a higher temporal resolution (lower $R$) and bigger codebook size $K$, the audio quality improves. On the other hand, this produces longer sequences over bigger vocabularies, leading to a harder task for the language model downstream. 
In other words, to generate high-quality audio samples, one needs to work at lower compression rates which requires stronger auto regressive models for sequence modeling. We evaluate different temporal resolutions and vocabularies sizes and report results in Section~\ref{sec:abb}.

\section{Experiments}
\noindent {\bf Implementation Details.}
\label{sec:details}
The compression model $A$ is implemented as an encoder-decoder model with a vector-quantization bottleneck. The encoder-decoder modules are similar to the ones in \cite{kreuk2022audiogen}: 4 encoder blocks with a 1D-convolutional layer at the beginning and end. Each encoder block is comprised of 3 residual blocks and a convolutional layer for down-sampling the temporal axis. The decoder is the reverse of the encoder. We trained $A$ for 100k steps with a batch size of 176 and a learning rate of 3e-4 using the Adam optimizer. The model was trained on 8 A100 GPUs. 

The audio language model is a transformer-decoder implementation from the fairseq repo.\footnote{Specifically we use the \texttt{transformer\_lm\_gpt3\_medium} version.}. It contains 24 transformer layers with 16 attention heads, embedding dimension of 1024, FFN size of 4096 and GELU activation functions. Each sample contains 10 seconds of speech (160,000 samples), represented by $\frac{10000}{R}$ discrete units (e.g., 5000 discrete units for $R=2$ms). We trained the audio language model until convergence with a batch size of 128 and a learning rate of 3e-4 using the Adam optimizer and an inverse square-root decay schedule and 3000 warm-up steps using 32 A100 GPUs.

\noindent {\bf Dataset.}
\label{sec:dataset}
We use the Audioset dataset \cite{gemmeke2017audio} for training and evaluating our model. Audioset consists of 2.1 million samples, or 5.8k hours of video, gathered from Youtube. The samples are categorized using an anthology of 527 classes with a wide variety such as human sounds (speech, whistling, respiratory), music (drums, guitar), things (engine, alarm, fire), etc. For the purposes of audio modeling, we extract the audio track from the collection of videos. For training and evaluation we use the standard split of \texttt{unbalanced\_train\_segments} for training and \texttt{eval\_segments} for evaluations.

\noindent {\bf Baselines.}
\label{sec:baselines}
We compared the proposed method to a set of existing methods for encoding audio into discrete representations. It was shown in \cite{lakhotia2021generative, kharitonov2021text, polyak2021speech} that the combination of a pre-trained SSL model with k-means clustering produces discrete audio representations suitable for speech synthesis. The resulting discrete representations were used to train a set of audio language models as baselines. For representations, we explore the use of contrastive predictive coding (CPC)~\cite{schneider2019wav2vec, riviere2020unsupervised}, wav2vec2.0~\cite{baevski2020wav2vec} as feature extractors, followed by a k-means clustering step. We start by training CPC and wav2vec2.0 on audioset, then train a k-means model on the learned representations. We set the number of clusters for the k-means procedure (and the resulting vocabulary size) to be 2048. We also explore log-melspectrograms as an additional naive baseline, in this case only the k-means step is required.
To reconstruct the time-domain signal from the above representations we train a variation of the HiFi-GAN neural vocoder proposed in \cite{polyak2021speech,kharitonov2022textless}. We trained CPC and wav2vec2.0 for 400k update steps using the official recipes.

\begin{table}[t!]
\centering
\caption{A comparison of the proposed method against existing baselines. The audio quality if measured using the MOS, continuation quality is measured using the cMOS. Results are reported for several time-resolutions. We report mean scores together with CI95 values. All reported results were obtained using codebook size of 2048.}
\label{tab:results}
\resizebox{0.8\columnwidth}{!}{
\begin{tabular}{l|c|c}
Model                        &  MOS             & cMOS \\
\toprule
Ground-truth                 &  3.51$\pm$0.29   & 3.54$\pm$0.38 \\
\midrule
Log-mel@10ms                 &  2.63$\pm$0.19   & 2.91$\pm$0.38 \\
Log-mel@2ms                  &  2.49$\pm$0.30   & 2.86$\pm$0.18 \\
CPC@10ms                 &  2.61$\pm$0.36   & 3.05$\pm$0.31 \\
wav2vec2.0@20ms              &  2.48$\pm$0.37   & 3.02$\pm$0.36 \\
\midrule
\audiogen@8ms                     &  2.82$\pm$0.25   & 3.33$\pm$0.28 \\
\audiogen@4ms                     &  2.87$\pm$0.28   & 3.23$\pm$0.29 \\
\audiogen@2ms                     &  3.03$\pm$0.29   & 3.18$\pm$0.26 \\
\bottomrule
\end{tabular}}
\vspace{-0.2cm}
\end{table}

\begin{figure*}
     \centering
     \begin{subfigure}[b]{0.28\textwidth}
         \centering
         \includegraphics[width=\textwidth]{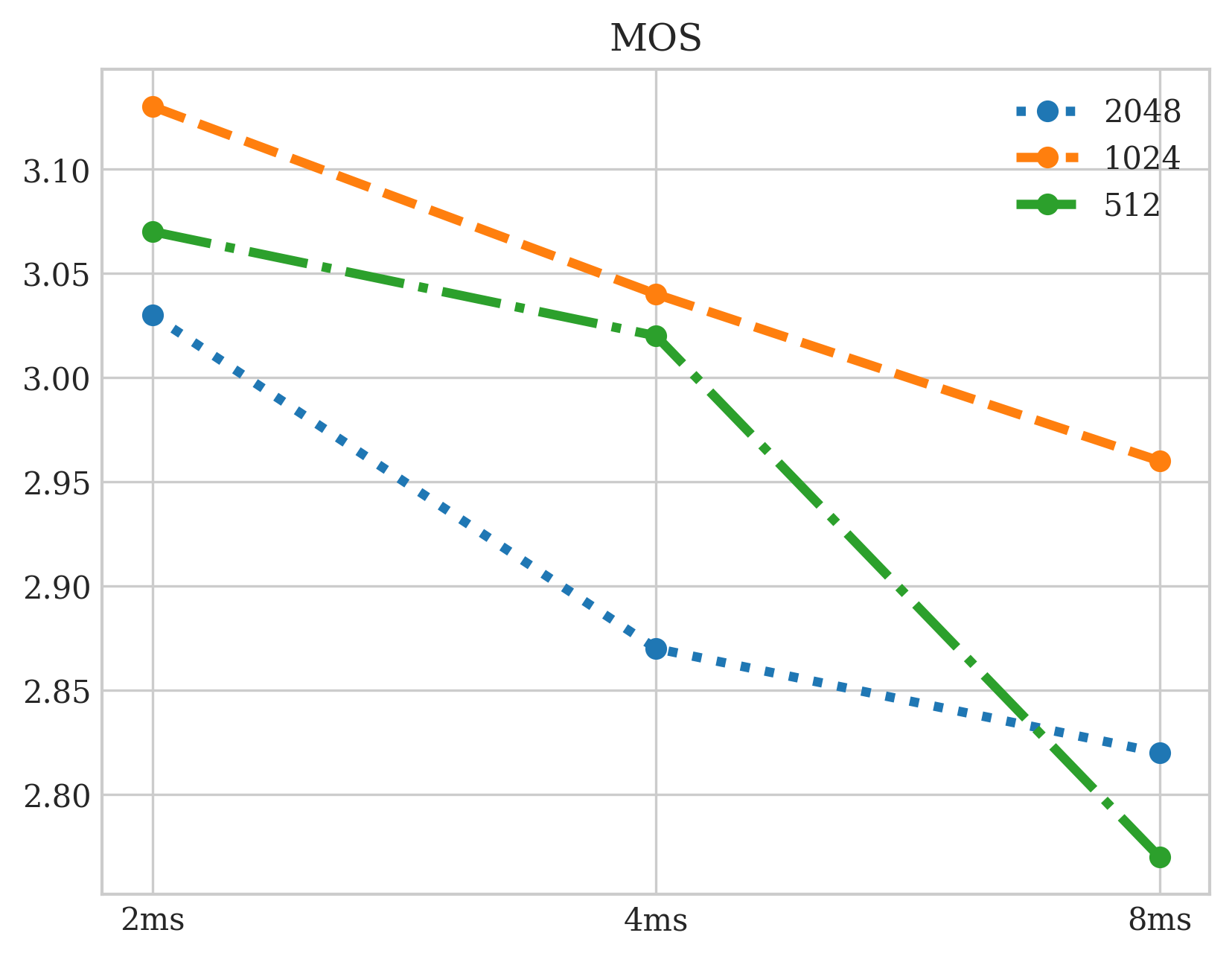}
         \caption{}
         \label{fig:mos}
     \end{subfigure}
     \hfill
     \begin{subfigure}[b]{0.28\textwidth}
         \centering
         \includegraphics[width=\textwidth]{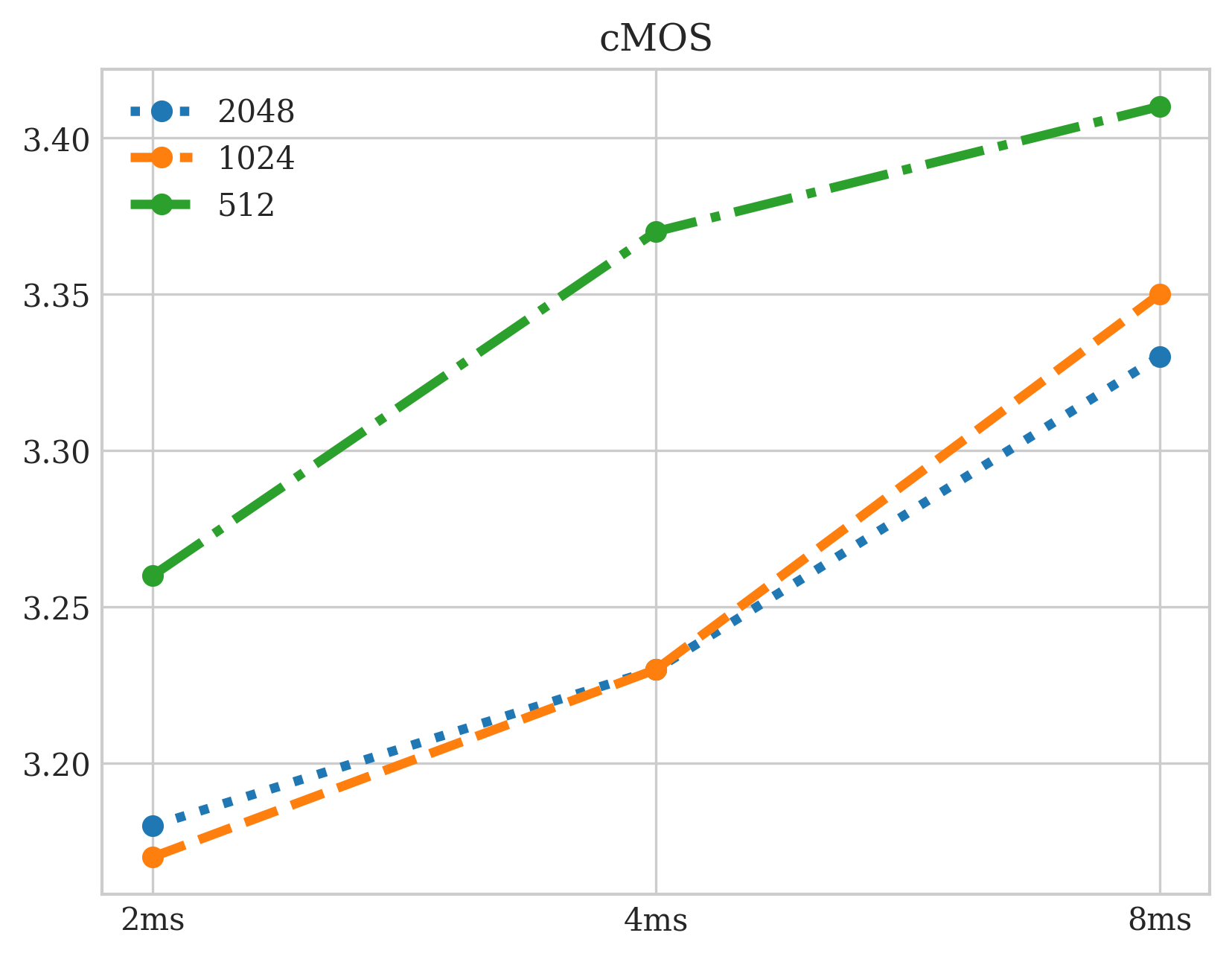}
         \caption{}
         \label{fig:cmos}
     \end{subfigure}
     \hfill
     \begin{subfigure}[b]{0.28\textwidth}
         \centering
         \includegraphics[width=\textwidth]{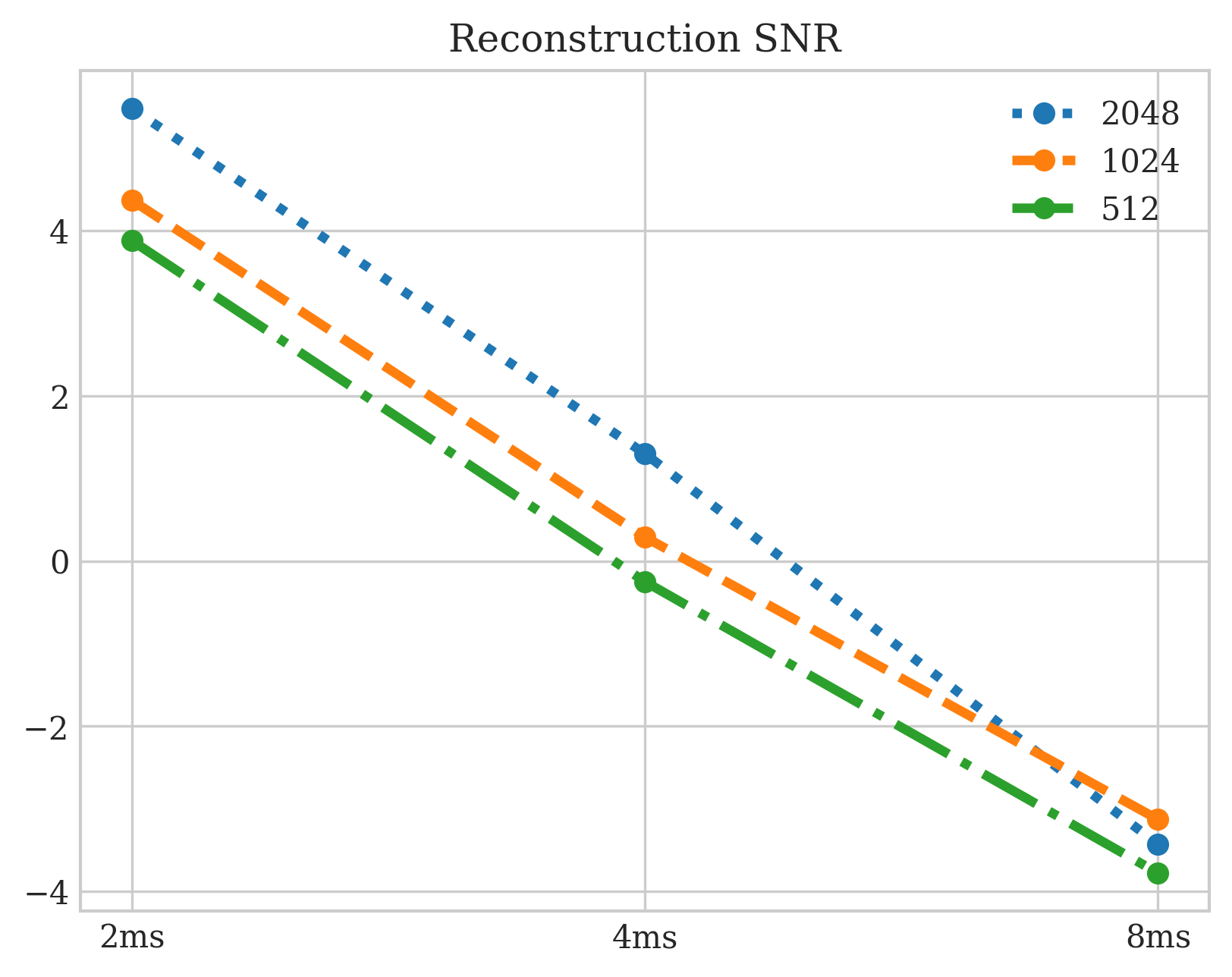}
         \caption{}
         \label{fig:snr}
     \end{subfigure}
    \caption{MOS and cMOS results for the proposed method as a function of the down-sampling factor. The x-axis presents the receptive filed corresponds to a single discrete unit. We additionally report reconstruction SNR between the target sample to the synthesized one without performing audio continuation. Results are reported for three codebook sizes: \{512, 1024, 2048\}.}
    \vspace{-0.2cm}
    \label{fig:abb}
\end{figure*}

\noindent {\bf Metrics.}
\label{sec:metrics}
We use two subjective metrics for model evaluation. The first is the Mean-Opinion-Score (MOS) for measuring the perceived audio quality. While the model output can be perceived as of high-quality, it might not be a suitable continuation to the audio prompt (model input). We therefore propose to use a new score, namely continuation-MOS (cMOS). 

In a cMOS study a human listener is presented with and audio waveform divided into two segments by a ``beep'' sound in the middle. The audio in the first half is the ground-truth audio used to prompt the model, and the audio in the second half is generated by our model. As control, we also evaluate the case where the second half is also ground-truth audio. The listener is instructed to rate on a scale of 1 to 5 the extent to which the second half of the audio matches the first half. In other words, if the second half is a possible continuation of the first half, it should be ranked higher and vice-versa. All participants were recruited using the Amazon Mechanical Turk platform. The CrowdMOS package \cite{ribeiro2011crowdmos} was used in all subjective experiments using the recommended recipes for outlier removal. All participants are native English speakers.

\noindent {\bf Results.}
\label{sec:results}
Table \ref{tab:results} summarizes the results for the proposed method and evaluated baselines. For a fair comparison, we tune the down-sampling factor of proposed method to reach temporal resolution of 8ms. Results suggest that the proposed method is superior to the evaluated baselines. 

Throughout preliminary experiments, we notice that the temporal resolution greatly affects the generated audio quality. With CPC and log-melspectrograms operating at a resolution of 10ms and wav2vec2.0 operating at 20ms their perceived audio quality is noticeably lower than that of the proposed method. For a complete comparison, we additionally train a model based on log-melspectrograms with smaller hop-length and window size, resulting in a higher temporal resolution of 2ms. We additionally report results for the proposed method using different temporal resolutions, specifically we report results for 8ms, 4ms, and 2ms temporal resolution (lower part of Table~\ref{tab:results}). As expected, as we increase the temporal resolution (lower $R$), results are improving. Interestingly, we observe the opposite trend when considering Log-mel features. When considering audio quality, results suggest that operating at a higher temporal resolution (lower $R$) is more beneficial. Conversely, when considering the continuation quality, results suggest that lower temporal resolution (higher $R$) produces shorter sequences, making them easier to model, and results in better cMOS.

Recall, the sampling time of the method in~\cite{dhariwal2020jukebox} was estimated at 8 hours for generating 1 minute of audio. Generating 1 minute of audio using the proposed method takes 25 minutes, demonstrating the efficiency of the proposed method.

\noindent {\bf Ablation study.}
\label{sec:abb}
Next, we conduct an ablation study to better understand the affect of temporal resolution and codebook size on the perceived audio quality and perceived continuation quality. Specifically, we evaluate a grid of \{2ms, 4ms, 8ms\} temporal resolutions and \{512, 1024, 2048\} codebook sizes. We compute both MOS, cMOS, and a Signal-to-Noise Ratio (SNR) between the target audio sample to the reconstructed one, denoted as reconstruction SNR. Notice, the reconstruction SNR does not involve any LM inference as we evaluate against the target signal. The premise behind this experiment is to solely evaluate the quality of the compression rate using different codebook sizes and temporal resolutions. Results are summarized in Figure~\ref{fig:abb}. 

Results suggest that the temporal resolution has a noticeable effect on both MOS and cMOS. When operating on a higher temporal resolution, the perceived quality is higher, this can be explained by working with a discrete representation with a higher sampling rate. Similarly to previous findings, having longer sequences results in a degradation in continuation quality. Interestingly, using a codebook size of 512 was shown to perform better than both 1024 and 2048 considering cMOS, while for for MOS 1024 yields the best overall performance. Regarding signal reconstruction, as expected increasing both temporal resolution and codebook size results with higher SNR.

\vspace{-0.2cm}
\section{Conclusion \& Future Work}
\vspace{-0.1cm}

In this work, we formalize the task of \emph{Audio Language Modeling}, where our goal is to learn the acoustic characteristics of general audio from the raw waveform. Using such a model, given an input audio prompt we first encode the signal into a discrete representation, sample from the language model to perform conditional audio generation, and synthesize the audio back to the time-domain. We study the effect of different audio encoders and time resolution on the generated audio samples. We perform an ablation study and discuss the trade-offs between generation quality and sequential modeling. We find that the perceived audio quality is greatly affected by the temporal resolution of the discrete encoding. However, working at higher temporal-resolution results in degradation of consistency over long sequences.
Evaluating such models is a subjective task by nature. For future work, we would like to design an improved evaluation procedure and include objective metrics to facilitate faster and less costly research.

\bibliographystyle{IEEEbib}
\bibliography{mybib}
\end{document}